\documentclass[sigplan,10pt]{acmart}
\renewcommand\footnotetextcopyrightpermission[1]{}
\setcopyright{none}
\acmConference[]{}{}{}

\newcommand{\citey}[1]{\cite{#1}}

\usepackage{amsmath}
\usepackage{url}
\usepackage{graphicx}
\usepackage{amsmath}
\usepackage{amssymb}
\usepackage{listings}
\usepackage{varwidth}
\usepackage{framed}
\usepackage{float}
\usepackage{fancyvrb}
\usepackage{proof}
\usepackage{natbib}

\usepackage{tikz}
\usetikzlibrary{decorations.text,calc,fit,arrows,chains,matrix,positioning,scopes,shapes.misc}

\newcommand{\tikzmark}[1]{%
     \tikz[overlay,remember picture] \node[inner sep=2pt] (#1) {};}

\newcommand{\ShiftForArrows}{0.0em,+0.5ex}%
\newcommand{\DrawArrows}[5][]{%
    \coordinate (Start) at ($(#2) + (\ShiftForArrows)$);
    \coordinate (End) at ($(#3) + (\ShiftForArrows)$);

    \draw[-stealth, thick, #1] (Start) to node[text=black,#5] {#4} (End);
}

\newcommand{\DrawBox}[3][]{%
    \draw [#1] ($(#2)-(0.1em,0.25ex)$) rectangle ($(#3)+(0.1em,1.5ex)$);
}
\makeatletter

\def\tikz@lib@parse@join#1{%
  \def\tikz@temp{#1}%
  \ifx\tikz@temp\pgfutil@empty%
    \tikz@lib@join@by@nodes by {} nodes \pgfutil@stop%
  \else%
    \pgfutil@in@{nodes }{#1}%
    \ifpgfutil@in@%
      \tikz@lib@parse@join@nodes#1\pgfutil@stop%
    \else%
      \tikz@lib@parse@join@nodes#1 nodes \pgfutil@stop%
    \fi%
  \fi%
}

\def\tikz@lib@parse@join@nodes#1nodes #2\pgfutil@stop{%
  \pgfutil@in@{by }{#1}%
  \ifpgfutil@in@%
    \tikz@lib@parse@join@by#1nodes #2\pgfutil@stop%
  \else%
    \tikz@lib@parse@join@by#1by {} nodes #2\pgfutil@stop%
  \fi%
}

\def\tikz@lib@parse@join@by#1by #2 nodes #3\pgfutil@stop{%
  \pgfutil@in@{with }{#1}%
  \ifpgfutil@in@%
    \tikz@lib@join@with@by@nodes#1by #2 nodes #3\pgfutil@stop%
  \else%
    \tikz@lib@join@by@nodes by #2 nodes #3\pgfutil@stop%
  \fi%
}

\def\tikz@lib@join@with@by@nodes with #1 by #2 nodes #3\pgfutil@stop{%
  \tikzset{after node path={(#1)edge[every join,#2]#3(\tikzchaincurrent)}}%
}

\def\tikz@lib@join@by@nodes by #1 nodes #2\pgfutil@stop{%
  \tikzset{after node path={%
             \ifx\tikzchainprevious\pgfutil@empty%
             \else%
               (\tikzchainprevious)edge[every join,#1]#2(\tikzchaincurrent)%
             \fi}}%
}

\makeatother

\newcommand{\incggen}[2][]{\includegraphics[#1]{latex.out/#2}}

\definecolor{lightgray}{rgb}{.9,.9,.9}
\definecolor{darkgray}{rgb}{.4,.4,.4}
\definecolor{purple}{rgb}{0.65, 0.12, 0.82}
\definecolor{mygreen}{rgb}{0,0.6,0}

\definecolor{Xcolor}{rgb}{0,0,1}
\definecolor{Vcolor}{rgb}{1,0,0}
\definecolor{Bcolor}{rgb}{0,1,0}

\lstnewenvironment{rmclisting}
{
\lstset{
   language=C++,
   keywordstyle={\color{black}},
   emphstyle={\color{red}},
   emph={L,LE,LR,XEDGE,VEDGE,PEDGE,XEDGE_HERE,VEDGE_HERE},
   emphstyle={[2]\color{mygreen}},
   emph={[2]wdata,wflag,rflag,rdata,echeck,insert,eupdate,dcheck,read,dupdate,load,store,pre,post,label,write1,push1,read1,write2,push2,read2,trylock,unlock,before,after,acquired,wa,wb,wc,wd,ra,r,init,update,lookup,get_widget_lookup,publish},
   extendedchars=true,
   basicstyle=\footnotesize\ttfamily,
   escapeinside={(*@}{@*)},
   columns=fullflexible,
}
}{}
\newcommand{\pushe}{\arr{pu}}

\mathchardef\col="003A  % \col for binding colon (mathcode ordinary: less space)
\mathchardef\semi="603B % \semi for (regular) semicolon
\mathchardef\bang="6021 % \bang for exclamation point
\mathchardef\lt="313C  % \lt for <
\mathchardef\gt="313E  % \gt for >
\newcommand{\bigleft}{\left(\begin{array}{@{}l@{}}}
\newcommand{\bigright}{\end{array}\right)}

\makeatletter
\def\blfootnote{\xdef\@thefnmark{}\@footnotetext}
\makeatother

\chardef\tilde="7E
\chardef\caret="5E
\newcommand{\arr}[1]{\stackrel{\sf #1}{\rightarrow}}

\newcommand{\rf}{\arr{rf}}

\newcommand{\vo}{\arr{vo}}

\newcommand{\xo}{\arr{xo}}

\newcommand{\integrityaggregateSlowdown}{1.05}

\newcommand{\weaknessaggregateSlowdown}{1.50}

\newcommand{\yeezyrculistSlowdown}{1.03}

\newcommand{\yeezyqspinlockSlowdown}{5.19}

\newcommand{\yeezyaggregateSlowdown}{1.65}

\newcommand{\li}{\lstinline}
\newcommand{\ttt}{\texttt}

\begin{document}

\setlength{\pdfpageheight}{\paperheight}
\setlength{\pdfpagewidth}{\paperwidth}

%\conferenceinfo{CONF 'yy}{Month d--d, 20yy, City, ST, Country}
%\copyrightyear{20yy}
%\copyrightdata{978-1-nnnn-nnnn-n/yy/mm}
%\copyrightdoi{nnnnnnn.nnnnnnn}

% Uncomment the publication rights you want to use.
%\publicationrights{transferred}
%\publicationrights{licensed}     % this is the default
%\publicationrights{author-pays}

%\titlebanner{Implementation of RMC}        % These are ignored unless
%\preprintfooter{Implementation of RMC}   % 'preprint' option specified.

%\title{Implementing a Calculus for Relaxed Memory}
\title{Compiling a Calculus for Relaxed Memory}
\subtitle{Practical constraint-based low-level concurrency \\ November 2017}

%\acmConference[]{}{}{}
%\acmYear{2017}

\author{Michael J. Sullivan}
\affiliation{\institution{Carnegie Mellon University}}
\email{mjsulliv@cs.cmu.edu}

\author{Karl Crary}
\affiliation{\institution{Carnegie Mellon University}}
\email{crary@cs.cmu.edu}
\author{Salil Joshi}
\affiliation{\institution{AlphaGrep Securities}}
\email{salil.ssj@gmail.com}

\ifdefined\ANON
\newcommand{\citeComp}{\cite{anon:rmc-compiler}}
\else
\newcommand{\citeComp}{\cite{sully:rmc-compiler}}
\fi

\begin{abstract}
Crary and Sullivan's Relaxed Memory Calculus (RMC) proposed a new
declarative approach for writing low-level shared memory concurrent
programs in the presence of modern relaxed-memory multi-processor
architectures and optimizing compilers.
In RMC, the programmer explicitly specifies
constraints on the order of execution of operations and on the
visibility of memory writes. These constraints are then enforced by
the compiler, which has a wide degree of latitude in how to accomplish
its goals.

We present \texttt{rmc-compiler}, a Clang and LLVM-based compiler for
RMC-extended C and C++. In addition to using barriers to enforce
ordering, \texttt{rmc-compiler} can take advantage of control and data
dependencies, something that is beyond the abilities of current C/C++
compilers. In \texttt{rmc-compiler}, RMC compilation is modeled
as an SMT problem with a cost term; the solution with the minimum cost
determines the compilation strategy.
In testing on ARM and POWER devices, RMC performs quite well, with
modest performance improvements relative to C++11 on most of our data
structure benchmarks and (on some architectures) dramatic improvements
on a read-mostly list test that heavily benefits from use of data
dependencies for ordering.

\end{abstract}

\maketitle

%\category{CR-number}{subcategory}{third-level}
%\keywordskeyword1, keyword2

\section{Introduction}

Writing programs with shared memory concurrency is notoriously
difficult even under the best of circumstances. By ``the best of
circumstances'', we mean something specific: when memory accesses are
sequentially consistent. Sequential consistency promises that threads
can be viewed as strictly interleaving accesses to a single shared
memory \cite{lamport:sequential-consistency}.
Sequential consistency is easy to understand and supports
some straightforward reasoning principles, though the exponential
growth of the state-space of concurrent programs leaves things still
quite tricky.

Modern multi-processor architectures, however, do not provide
sequential consistency. Many processors execute instructions
out-of-order that are observable from other processors while
preserving the behavior of single-threaded computations. More
frighteningly, memory subsystems with hierarchical caches and store
buffers can themselves propagate writes out of order and even to
different processors in different orders. Not to be outdone, compilers
get in on the sequential-consistency-violating fun as well: many
transformations that are perfectly sensible in a single-threaded
setting violate sequential consistency. Excitingly, this includes some
bread-and-butter optimizations like common subexpression elimination
and loop invariant code motion.

Because the compiler is intimately involved in the problem and because
it is preferable to abstract away from the differences between
architectures, this is a language design issue \cite{boehm:threads-not-library}.
The now-standard language approach to this problem, then, is for
languages to guarantee that
data-race-free code will behave in a sequentially consistent
manner. Programmers can then use locks and other techniques to
synchronize between threads and rule out data races. This may not be
good enough, however, for
performance-critical code and library implementation,
requiring languages that target these domains to provide
a well defined low-level mechanism for shared memory concurrency. C
and C++ (since the C11 and C++11 standards) provide a mechanism based
around specifying ``memory orderings'' when accessing concurrently
modified locations \cite{iso:cpp14}.
These memory orderings induce constraints that
constrain the behavior of programs. While some fragments are fairly
pleasant to use, the system as a whole is quite complex and contains
many moving parts.

\citeauthor{crary+:rmc}'s Relaxed Memory Calculus (RMC) proposes a more
declarative new approach to handling weak memory in low-level
concurrent programming: explicit, programmer-specified constraints
\citey{crary+:rmc}. In RMC, the programmer explicitly specifies
constraints on the order of execution of operations and on the
visibility of memory writes. These constraints are then enforced by
the compiler, which has a great deal of latitude in how to achieve
them. Because of the very fine-grained information about permissible
behaviors, this can allow the generation of more efficient code.

We present a compiler for RMC.
The paper is organized as follows:
\begin{itemize}
\item In Section \ref{section:tutorial} we recapitulate the design of
  RMC in the form of a tutorial introduction to C++ programming using
  RMC with \texttt{rmc-compiler}.

\item We present a new informal model (Section \ref{section:model}) of
  the execution of RMC programs. We detail the split into an
  \emph{execution model} (Section \ref{section:execution}) that models
  the potentially out-of-order execution of program instructions and a
  \emph{memory system model} (Section \ref{section:coherence}) that
  determines what values are read by memory reads.

\item We discuss \texttt{rmc-compiler}, our LLVM-based compiler for
  RMC-extended languages and the compilation of RMC to x86, ARM, and
  POWER (Section \ref{section:compiler}).

\item We detail the potential and difficulties of optimizing barrier
  placement for RMC programs (Section \ref{section:optapproach}) and
  our modeling of the problem as an SMT problem (Section \ref{section:smt}).

\item We evaluate the performance of RMC programs (Section
  \ref{section:eval}).
\end{itemize}

\section{Programming with RMC}
\label{section:tutorial}
%Some of this discussion is adapted from \citet{crary+:rmc}.
% XXX: do I need to cite my thesis??

%\subsection{Basics}

The Relaxed Memory Calculus (RMC) is a declarative approach to
handling memory ordering in low-level lock-free concurrent
programming. In RMC, the programmer can explicitly and directly
specify the key ordering relations that govern the behavior of the
program.

These key relations---which we will also refer to as ``edges''---are
that of {\em visibility-order\/} ($\vo$) and {\em execution-order\/} ($\xo$).
To see the intended meaning of these relations, consider this pair of
simple functions for passing a message between two threads:

\begin{minipage}[b]{0.40\linewidth}
\begin{rmclisting}
int data, (*@\tikzmark{flagL}@*)flag(*@\tikzmark{flag}@*);

void send(int msg) {
  (*@\tikzmark{wdataL}@*)data = msg;(*@\tikzmark{wdata}@*)
  (*@\tikzmark{wflagL}@*)flag = 1;(*@\tikzmark{wflag}@*)
}
\end{rmclisting}
\end{minipage}
\begin{minipage}[b]{0.15\linewidth}
\end{minipage}
\begin{minipage}[b]{0.3\linewidth}
\begin{rmclisting}
int recv() {
  while (!(*@\tikzmark{rflagL}@*)flag(*@\tikzmark{rflag}@*))
    continue;
  return (*@\tikzmark{rdataL}@*)data(*@\tikzmark{rdata}@*);
}
\end{rmclisting}
\end{minipage}
\begin{tikzpicture}[overlay,remember picture]
\DrawBox[Bcolor]{flagL}{flag}

\DrawArrows[Vcolor, out=-20, in=10, distance=4.0ex]{wdata}{wflag}{vo}{right}
\DrawBox[Bcolor]{wdataL}{wdata}
\DrawBox[Bcolor]{wflagL}{wflag}

\DrawArrows[Xcolor, out=-20, in=50, distance=3.5ex]{rflag}{rdata}{xo}{right}
\DrawBox[Bcolor]{rdataL}{rdata}
\DrawBox[Bcolor]{rflagL}{rflag}
\end{tikzpicture}

The visibility edge ($\vo$) between the
writes in \li{send} ensures that the write to \li{data} is visible to
other threads before the write to \li{flag} is. Somewhat more
precisely, it means that any thread that can see the write to
\li{flag} can also see the write to \li{data}. The execution edge
($\xo$) between the reads in \li{recv} ensures that the reads from
\li{flag} occur before the read from \li{data} does.
%
%XXX: we \emph{visible to} the /third/ time it is used in this paragraph
% (but the first time that it is used in a technical sense).
This combination
of constraints ensures the desired behavior: the loop that reads
\li{flag} can not exit until it sees the write to \li{flag} in
\li{send}; since the write to \li{data} must become visible to a
thread first, it must be visible to the \li{recv} thread when it sees
the write to \li{flag}; and then, since the read from \li{data} must
execute after that, the write to \li{data} must be \emph{visible to}
the read.

%% \begin{figure}
%% \centering
%% \begin{minipage}[b]{0.40\linewidth}
%% \begin{rmclisting}
%% int data, flag;

%% void send(int msg) {
%%   data = msg;
%%   flag = 1;
%% }
%% \end{rmclisting}
%% \end{minipage}
%% \begin{minipage}[b]{0.15\linewidth}
%% \end{minipage}
%% \begin{minipage}[b]{0.3\linewidth}
%% \begin{rmclisting}
%% int recv() {
%%   while (!flag)
%%     continue;
%%   return data;
%% }
%% \end{rmclisting}
%% \end{minipage}
%% \caption{Message passing: take 1}
%% \label{fig:mp_broke}
%% \end{figure}

%%%%%%%%%%%%%%%%%%%%%%%%%%%%%%%%%%%%%%%%%%%%%%%%%%%%%%%%%%%%%

We can demonstrate this diagrammatically as a graph
of memory actions with the constraints as labeled edges:

%% \begin{center}
%% \begin{tikzpicture}[remember picture,every join/.append style={-stealth}]
%%   \matrix (m) [matrix of nodes, row sep=3em, column sep=3em]
%%     { W[data]=msg & \, \\
%%       W[flag]=1 & R[flag]=1 \\
%%       \,          & R[data]=msg \\
%%     };
%%   { [start chain] \chainin (m-1-1);
%%     \chainin (m-2-1) [join=by {thick,Vcolor} nodes {node[text=black,right] {vo}}]; }
%%   { [start chain] \chainin (m-2-2);
%%     \chainin (m-3-2) [join=by {thick,Xcolor} nodes {node[text=black,right] {xo}}]; }
%%   { [start chain] \chainin (m-2-1);
%%     \chainin (m-2-2) [join=by {thick,Vcolor,dashed,out=0,in=180} nodes
%%       {node[text=black,above] {rf}}]; }
%% % XXX: do we include this?
%%  {
%%     [start chain] \chainin (m-1-1);
%%     \chainin (m-3-2) [join=by {thick,Vcolor,dashed,out=210,in=180,distance=3cm} nodes {node[text=black,left] {rf}}]; }
%% \end{tikzpicture}
%% \end{center}

\begin{center}
\incggen[width=4cm]{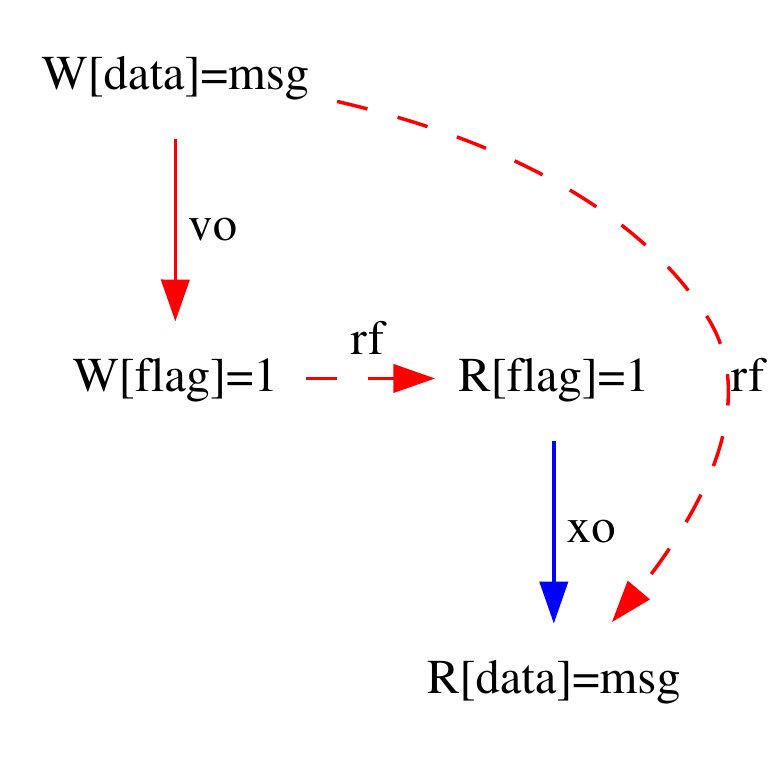}
\end{center}

In the diagram, the programmer specified edges ($\vo$ and $\xo$) are
drawn as solid lines while the ``reads-from'' edges (written $\rf$),
which arise dynamically at runtime, are drawn as dashed lines.
Since reading from a write is clearly a demonstration that the write
is visible to the read, we draw reads-from edges in the same color red
as we draw specified visibility-order edges, to emphasize that both
carry visibility.
Then, the chain of red visibility edges followed by the chain of
blue execution order edges means that the write to \li{data} is
visible to the read.

\subsection{Concrete syntax: tagging}

Unfortunately, we can't actually just draw arrows between
expressions in our source code, and so we need a way to describe these
constraints in text. We do this by tagging expressions with names and
then declaring constraints between tags:

%\begin{center}
\begin{minipage}[b]{0.45\linewidth}
\begin{rmclisting}
int data;
rmc::atomic<int> flag;

void send(int msg) {
  VEDGE(wdata, wflag);
  L(wdata, data = msg);
  L(wflag, flag = 1);
}
\end{rmclisting}
\end{minipage}
\begin{minipage}[b]{0.15\linewidth}
\end{minipage}
\begin{minipage}[b]{0.40\linewidth}
\begin{rmclisting}
int recv() {
  XEDGE(rflag, rdata);
  while (!L(rflag, flag))
    continue;
  return L(rdata, data);
}
\end{rmclisting}
\end{minipage}
%\end{center}

Here, the \li{L} construct is used to tag expressions. For example,
the write \li{data = msg} is tagged as \li{wdata} while the read from
\li{flag} is tagged \li{rflag}. The declaration
\li{VEDGE(wdata, wflag)} creates a visibility-order edge between
actions that are tagged \li{wdata} and actions tagged \li{wflag}.
\li{XEDGE(rflag, rdata)} similarly creates an execution-order edge.

Visibility order implies execution order, since it does not make sense
for an action to be visible before it has occured.

Visibility and execution edges only apply between actions in program
order. This is mainly relevant for actions that occur in loops, such
as:

%\begin{center}
\begin{minipage}[b]{0.3\linewidth}
\begin{rmclisting}
VEDGE(before, after);
for (i = 0; i < 2; i++) {
  L(after, x = i);
  L(before, y = i + 10);
}
\end{rmclisting}
\end{minipage}
%\end{center}

This generates visibility edges from writes to \li{y} to writes to
\li{x} in future iterations, as shown in this trace (in which
unlabeled black lines represent program order):

\incggen[scale=0.55]{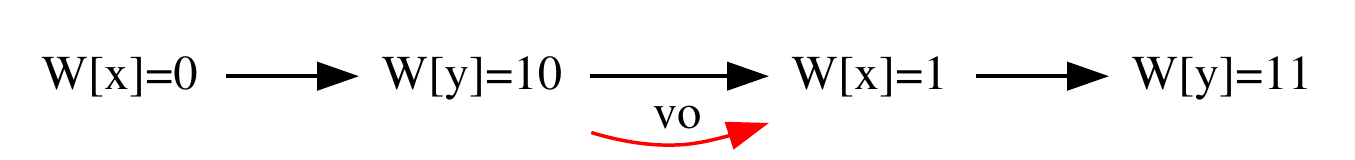}

Furthermore, edge declarations generate constraints between all
actions that match the tags, not only the ``next'' one. If we flip the
\li{before} and \li{after} tags in the previous example, we get:

\begin{minipage}[b]{0.3\linewidth}
\begin{rmclisting}
VEDGE(before, after);
for (i = 0; i < 2; i++) {
  L(before, x = i);
  L(after, y = i + 10);
}
\end{rmclisting}
\end{minipage}

\noindent which yields the following trace:

\incggen[scale=0.55]{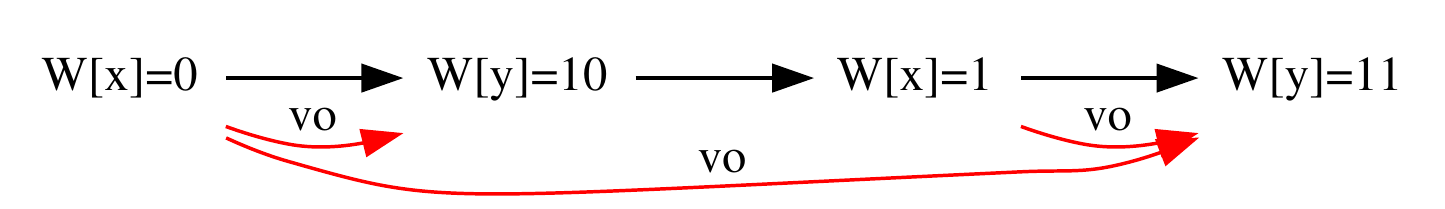}

In addition to the obvious visibility edges between writes in the same
loop iteration, we also have an edge from the write to \li{x} in the first
iteration to the write to \li{y} in the second. This behavior will be
important in the ring buffer example in Section \ref{section:ringbuf}.

%\XXX{This is kind of janky; I want a better explanation}
While this behavior is a good default, it is sometimes necessary to
have more fine-grained control over which matching actions are
constrained. This can be done with ``scoped'' constraints:
\li{VEDGE_HERE(a, b)} establishes visibility edges between executions
of \li{a} and \li{b}, but only ones that do not leave the ``scope'' of
the constraint\footnote{ Where the ``scope'' of a constraint is
  defined (somewhat unusually) as everything that is dominated by the
  constraint declaration in the control flow graph.}.  We can modify
the above example with a scoped constraint:

\begin{minipage}[b]{0.3\linewidth}
\begin{rmclisting}
for (i = 0; i < 2; i++) {
  VEDGE_HERE(before, after);
  L(before, x = i);
  L(after, y = i + 10);
}
\end{rmclisting}
\end{minipage}

\noindent which yields the following trace in which the edges between
iterations of the loop are not present:

\incggen[scale=0.55]{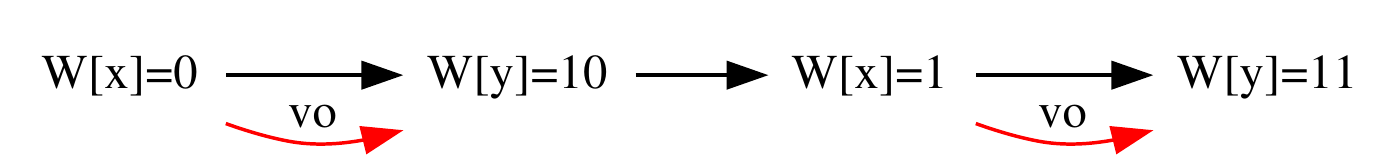}

%%%%%%%%%%%%%%%%%%%%%%%%%%%%%%%%%%%%%%%%%%%%%%%%%%%%%%%%%%%%%

\subsection{Pre and post edges}
\label{section:pre_post}

We have discussed drawing fine-grained constraint edges
between actions. Sometimes, however, it is necessary to declare
visibility and execution constraints in a much more coarse-grained
manner. This is particularly common at library module boundaries,
where it would be unwieldy and abstraction
breaking to need to
specify fine-grained edges between a library and client code.
To accommodate these needs, RMC supports special \li{pre} and \li{post}
labels that allow creating edges between an action and \emph{all} of
its program order predecessors or successors.

One of
the most straightforward places where coarse-grained constraints are
needed are in the implementation of locks. Here, any actions performed
during the critical section must be visible to any thread that has
observed the unlock at the end of it, as well as not being executed
until the lock has actually been obtained. This corresponds to the
actual release of a lock being visibility-order \emph{after}
everything before it in program order and the acquisition of a
lock being execution-order \emph{before} all of its program order
successors.

% XXX: phrasing
In this implementation of simple spinlocks, we do this with
post-execution edges from the exchange that attempts to acquire
the lock and with pre-visibility edges to the write that releases the
lock:

\begin{minipage}[b]{0.50\linewidth}
\begin{rmclisting}
void spinlock_lock(spinlock_t *lock) {
  XEDGE(trylock, post);
  while (L(trylock, lock->state.exchange(1)) == 1)
    continue;
}
\end{rmclisting}
\end{minipage}

\begin{minipage}[b]{0.40\linewidth}
\begin{rmclisting}
void spinlock_unlock(spinlock_t *lock) {
  VEDGE(pre, unlock);
  L(unlock, lock->state = 0);
}
\end{rmclisting}
\vspace{1.25em}
\end{minipage}

Pre and post edges are a critical piece of the RMC design:
the standard pattern for implementing concurrency objects in RMC is to
use point-to-point edges for internal constraints and pre/post edges
for interfacing with client code.

\subsection{Transitivity}

Visibility order and execution order are both transitive. This means
that, although the primary meaning of visibility order is in
how it relates writes, it
is still useful to create edges between other sorts of actions.

% DITCHED
%% In fact, because of this transitivity, it is even sometimes profitable
%% to create edges to no-ops! In the \li{spinlock_lock} example before,
%% we draw an execution edge from \li{trylock} to the quasi-tag
%% \li{post}. This means that each exchange on the lock is execution
%% ordered before not only the body of the critical section, but, if the
%% test-and-set fails, any future test-and-set attempts. This is stronger
%% than is actually necessary, and we can weaken it using a no-op:

%% \begin{minipage}[b]{0.80\linewidth}
%% \begin{rmclisting}
%% void spinlock_lock(spinlock_t *lock) {
%%   XEDGE(trylock, acquired);
%%   XEDGE(acquired, post);
%%   while (L(trylock, lock->state.exchange(1)) == 1)
%%     continue;
%%   L(acquired, noop());
%% }
%% \end{rmclisting}
%% \end{minipage}

%% Here, the exchange is specified to execute before \li{acquired},
%% which is a no-op that is specified to execute before all of its
%% successors. Since a no-op doesn't \emph{do} anything, nothing is
%% needed to ensure the execution order with the no-op, but transitivity
%% ensures that the lock attempts are execution ordered before everything
%% after \li{acquired}.

%A somewhat more substantive and less niche application of
%transitivity of visibility occurs with visibility edges from reads to
%writes.
The primary substantive application of transitivity occurs
with visibility edges from reads to writes.
Consider the following trace, which shows a variation on
message passing (known as ``WWC''):
\begin{center}
\incggen[scale=0.55]{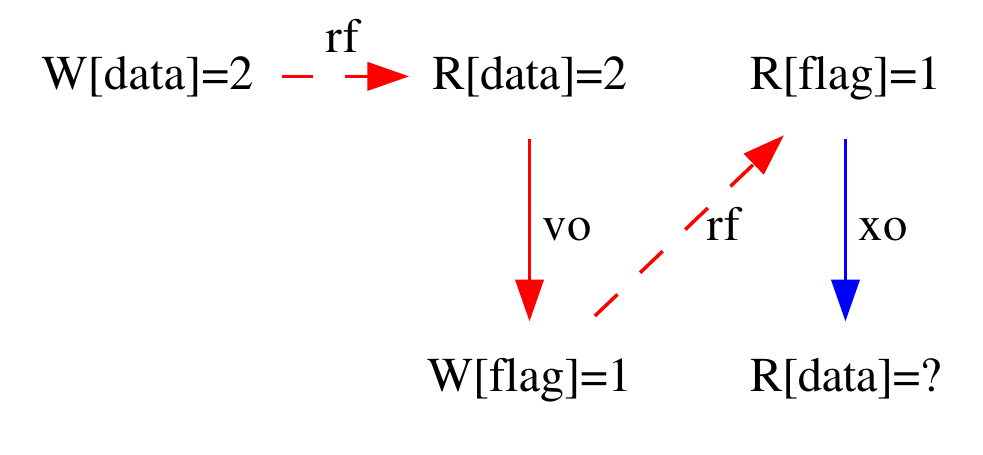}
\end{center}

Since reads-from is a form of visibility, and since visibility is
transitive, this means that \li{W[data]=2} is visible before
\li{W[flag]=1}. It is then also visibility ordered before
\li{R[flag]=1}; since that must execute before \li{R[data]=?},
this means that \li{W[data]=2} must be \emph{visible to}
\li{R[data]=?}, which will then read from it.

\subsection{Pushes}

Visibility order is a powerful tool for controlling the
\emph{relative} visibility of actions, but sometimes it is necessary
to worry about \emph{global} visibility. One case where this might be
useful is in preventing store buffering behavior:

\begin{center}
\incggen[scale=0.55]{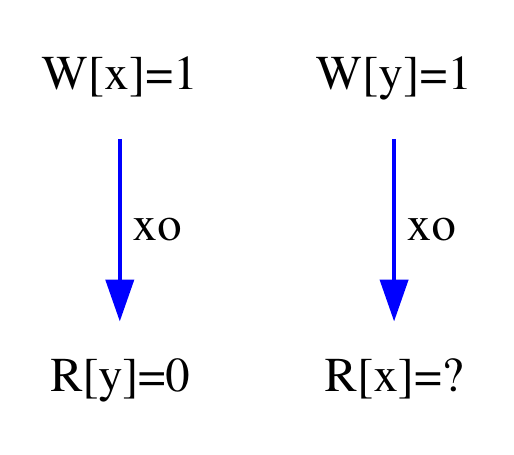}
\end{center}

Here, two threads each write a $1$ into a memory location and then
attempt to read the value from the other thread's location (this idiom
is the core of the classic ``Dekker's algorithm'' for two thread
mutual exclusion). In this trace, \li{R[y]=0} reads 0, and we would
like to require (as would be the case under sequential consistency)
that \li{R[x]=?} will then read $1$. However, it too can read $0$,
since nothing forces \li{W[x]=1} to be visible to it. Although there
is an execution edge from \li{W[x]=1} to \li{R[y]=0}, this only
requires that \li{W[x]=1} \emph{executes} first, not that it be
visible to other threads. Upgrading the execution edges to visibility
edges is similarly unhelpful; a visibility edge from a write to a read
is only useful for its transitive effects, and there are none
here. What we need is a way to specify that \li{W[x]=1} becomes
\emph{visible} before \li{R[y]=0} \emph{executes}.

Pushes provide a means to do this: when a push executes, it is
immediately globally visible (visible to all threads). As a
consequence of this, visibility between push operations forms a total
order. Using pushes, we can rewrite the above trace as:

\begin{center}
\incggen[scale=0.55]{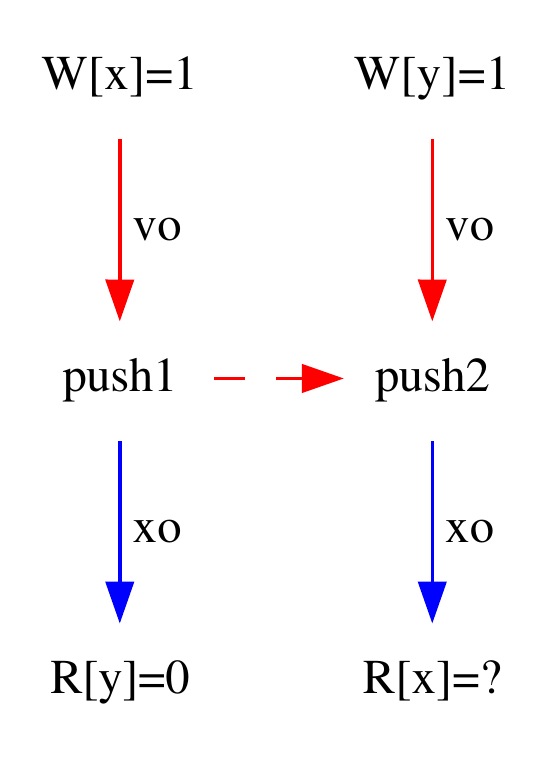}
\end{center}

Here, we have inserted a push that is visibility-after the writes and
execution-before the read. Since visibility among pushes is total,
either push1 or push2 is visible to the other. If push1 is visible before
push2, as in the diagram, then \li{W[x]=1} is visible to \li{R[x]=?},
which will then read $1$. If push2 was visible to push1, then
\li{R[y]=0} would be impossible, as it would be able to see the
\li{W[y]=1} write.

In the concrete syntax, the primary means of inserting a push is via a
``push edge'':

\begin{minipage}[b]{0.40\linewidth}
\begin{rmclisting}
  PEDGE(write1, read1);
  L(write1, x = 1);
  r1 = L(read1, y);
\end{rmclisting}
\end{minipage}

\noindent
A push edge from an action $a$ to $b$ means that a push will be
performed that is visibility after $a$ and execution before
$b$. Somewhat more informally, it means that $a$ will be globally
visible before $b$ executes.

\subsection{An example: ring-buffers}
\label{section:ringbuf}

As a realistic example of code using the RMC memory model, consider
the code in Figure~\ref{fig:ringbuffer}.  This code---adapted from the
Linux kernel~\cite{howells+:circular-buffers}---implements a
ring-buffer,
a common data structure that
implements an queue with a fixed maximum size.  The
ring-buffer maintains front and back pointers into an array, and the
current contents of the queue are those that lie between the back and
front pointers (wrapping around if necessary).  Elements are inserted
by advancing the back pointer, and removed by advancing the front
pointer.

% XXX: should this be a figure?
%\begin{figure}[H]
\begin{figure}
  %\begin{minipage}[b]{0.50\linewidth}

% goes between functions in a one-column version
%% \end{rmclisting}
%% \end{minipage}

%% \begin{minipage}[b]{0.50\linewidth}
%% \begin{rmclisting}

\begin{rmclisting}
bool buf_enqueue(ring_buf *buf, unsigned char c) {
  XEDGE(echeck, insert);
  VEDGE(insert, eupdate);

  unsigned back = buf->back;
  unsigned front = L(echeck, buf->front);

  bool enqueued = false;
  if (back - front < BUF_SIZE) {
    L(insert, buf->buf[back % BUF_SIZE] = c);
    L(eupdate, buf->back = back + 1);
    enqueued = true;
  }
  return enqueued;
}

int buf_dequeue(ring_buf *buf) {
  XEDGE(dcheck, read);
  XEDGE(read, dupdate);

  unsigned front = buf->front;
  unsigned back = L(dcheck, buf->back);

  int c = -1;
  if (back - front > 0) {
    c = L(read, buf->buf[front % BUF_SIZE]);
    L(dupdate, buf->front = front + 1);
  }
  return c;
}
\end{rmclisting}
%\vspace{1em}
%\end{minipage}
\caption{A ring buffer}
\label{fig:ringbuffer}
\end{figure}

This ring-buffer implementation is a single-producer, single-consumer,
lock-free ring-buffer. This means that only one reader and one writer
are allowed to access the buffer at a time, but the one reader and the
one writer may access the buffer concurrently.

In this implementation, we do not wrap the front and the back indexes
around when we increment them, but instead whenever we index into the
array. The number of elements in the buffer, then, can be calculated
as \li{back - front}.

There are two important properties we require of the ring-buffer: (1) the
elements dequeued are the same elements that we enqueued (that is,
threads do not read from an array location without the write to that
location being visible to it), and (2) no
enqueue overwrites an element that has not been dequeued

\newcommand{\dcheck}{{\text{\li{dcheck}}}}
\newcommand{\dread}{{\text{\li{read}}}}
\newcommand{\dupdate}{{\text{\li{dupdate}}}}
\newcommand{\echeck}{{\text{\li{echeck}}}}
\newcommand{\eupdate}{{\text{\li{eupdate}}}}
\newcommand{\ewrite}{{\text{\li{insert}}}}

The key lines of code are those tagged \echeck, \ewrite, and
\eupdate{} (in {\li{enqueue}}), and \dcheck, \dread, and \dupdate{} (in
{\li{dequeue}}).  (It is not necessary to use disjoint tag variables in
different functions; we do so to make the reasoning more clear.)

For property (1), the key constraints are $\ewrite \vo \eupdate$ and
$\dcheck \xo \dread$. If we consider an dequeue reading from some
enqueue, $\dcheck$ reads from $\eupdate$ and so
$\ewrite \vo \eupdate \rf \dcheck \xo \dread$. Thus $\ewrite$ is
visible to $\dread$. Note, however, that if there are more than one
element in the buffer, the $\eupdate$ that $\dcheck$ reads from will
not be the $\eupdate$ that was performed when this value was enqueued,
but one from some \emph{later} enqueue. That is just fine, and the
above reasoning still stands. As discussed above, constraints
apply to \emph{all} matching actions, even ones that do not occur
during the same function invocation. Thus the write of the value into
the buffer is visibility ordered before the \li{back} updates of all
future enqueues by that thread.

\begin{figure}
\centering
\incggen[scale=0.65]{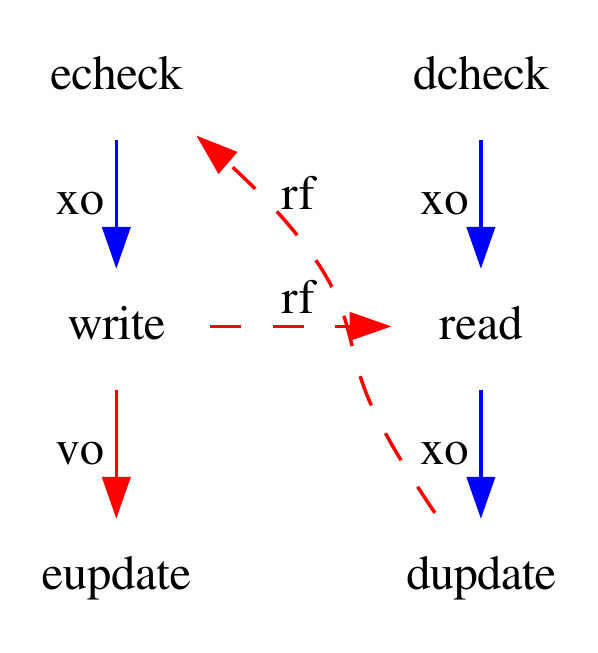}
\caption{Impossible ring-buffer trace}
\label{fig:ringbuffer-trace}
\end{figure}

Property (2) is a bit more complicated. % XXX: is it
The canonical trace we wish to prevent appears in
Figure~\ref{fig:ringbuffer-trace}.  In it, $\dread$ reads from
$\ewrite$, a ``later'' write that finds room in the buffer
only because of the space freed up by $\dupdate$. Hence,
a current entry is overwritten.

%XXX: should this be a figure?

This problematic trace is impossible, since $\dread \xo
\dupdate \rf
\echeck \xo \ewrite \rf \dread$.
Since you cannot read from a write that has not executed, writes must
be executed earlier than any read that reads from them. Thus this
implies that $\dread$ executes before itself, which is a contradiction.

%\vspace{-3em}
\subsection{Using data dependency}
%{\bf Using data dependency}
\label{section:datadeps}

One of the biggest complications of the C++11 model is the ``consume''
memory order, which establishes ordering based on what operations are
data dependent. This is useful because it allows read access to data
structures to avoid needing any barriers (on ARM and the like) in many
cases. This technique is extremely widespread in the Linux kernel.

\begin{figure}
%\begin{minipage}[b]{0.50\linewidth}
%goes between funcs in one-columns
%\end{rmclisting}
%\end{minipage}
%\begin{minipage}[b]{0.40\linewidth}
%\begin{rmclisting}
\begin{rmclisting}
rmc::atomic<widget *> widgets[NUM_WIDGETS];

void update_widget(char *key, int foo, int bar) {
  VEDGE(init, update);
  widget *w = L(init, new widget(foo, bar));

  int idx = calculate_idx(key);
  L(update, widgets[idx] = w);
}

// Some client code
int use_widget(char *key) {
  XEDGE_HERE(lookup, r);

  int idx = calculate_idx(key);
  widget *w =  L(lookup, widgets[idx]);
  return L(r, w->foo) + L(r, w->bar);
}
\end{rmclisting}
%\vspace{1em}
%\end{minipage}
\caption{A widget storing library}
\label{fig:widget}
\end{figure}

In Figure~\ref{fig:widget} we show a toy library for storing an array
of widgets that tries
to illustrate the shape of such code. In \li{update_widget}, a new
widget object is constructed and initialized and then a pointer is
written into an array; to ensure visibility of the initialization, a
visibility edge is used. In \li{use_widget}, which is a client
function to look up a widget and add together its two fields, the
message passing idiom is completed by execution edges from the lookup
to the actual accesses of the object.
The use of \li{XEDGE_HERE} is key to taking advantage of data
dependencies---data dependencies can't
enforce ordering with \emph{all} subsequent invocations of the
function, so we use \li{XEDGE_HERE} so that the ordering only needs
to apply within a given execution of the function.
The key thing about this code is
that it uses the same execution order idiom as message passing that
does not have data dependencies---in RMC, we just provide a uniform
execution order mechanism and rely on the compiler to be able to take
advantage of existing data dependencies in cases like this one.

In order to support this sort of idiom when the edges extend beyond
one function, RMC supports drawing edges that involve the "action" of
passing a value to a function or returning it as an argument, which
provides a way to specify fine-grained ordering constraints between
functions. The details of this are omitted for space reasons.

(Note that this code totally ignores the issue of freeing old widgets
if they are overwritten. This is a subtle issue; the solution
generally taken in Linux is the read-copy-update mechanism
\cite{mckenney+:rcu}.)

\section{Model Details}
\label{section:model}
We build on top of a draft version of RMC 2.0 \cite{crary+:rmc2},
which reworks the formalization of out-of-order execution and rules
out out-of-thin-air executions.

The model of how an RMC program is executed is split into two parts.
On one side, the \emph{execution model} models the potentially
out-of-order execution of actions in the program. On the other side,
the \emph{memory system model} determines what values are read by
memory reads.

Parts of the division of labor between the execution and memory system
side in RMC are somewhat unusual for an operational model: the
execution model is \emph{extremely} weak and relies on the memory
subsystem's coherence rules to enforce the correctness of
single-threaded code.

\subsection{Execution Model}
\label{section:execution}

In RMC, the responsibility of ensuring that single-threaded programs
behave as expected falls on the memory system model, which allows for
an extremely permissive execution model.
%% Intuitively, the model that we would like to have for how actions are
%% executed, is that actions may be executed in any order at all, except
%% as constrained by execution and visibility constraints.
Intuitively, our execution model is that actions may be executed in
any order at all, except as constrained by execution and visibility
constraints.
This includes
when the actions have a dependency between them, whether control or
data! That is, actions (both reads and writes) may be executed
speculatively under the assumption that some ``earlier'' read will
return a particular value.

%% Unfortunately, this extremely permissive model is \emph{too} weak, and
%% gives rise to problematic and bizarre behaviors known as ``thin-air
%% reads'' in which writing a speculated value is eventually used to
%% justify the read that produced it, in a sort of bizarre ``stable time
%% loop'' \footnote{Like the song Johnny B. Goode in ``Back to the Future''}.
%% While the original version of RMC allowed thin-air reads,
%% RMC 2.0 rules them out based on a
%% combination of ideas from the original RMC paper and the
%% ``Promising Semantics'' of \citet{kang+:promising}.
%% The heart of the idea is that before a thread can execute a write, it
%% must demonstrate that it could have executed the same write without
%% having any outstanding speculated reads. The exact details are subtle,
%% and outside the scope of this paper.
%% For most intuitive purposes, the hand-wavy prose rule in the C++14
%% standard does suffice: ``Implementations should ensure that no
%% ``out-of-thin-air'' values are computed that circularly depend on
%% their own computation'' \cite{iso:cpp14}.

\subsection{Memory system model}
\label{section:coherence}

The main question to be answered by a memory model is: for a given
read, what writes is it permitted to read from? Under sequential
consistency, the answer is ``the most recent'', but this is not a
necessarily a meaningful concept in relaxed memory settings.

While RMC does not have a useful \emph{global} total ordering of
actions, there does exist a useful \emph{per-location} ordering of
writes that is respected by reads. This order is called
\emph{coherence order}. It is a strict partial order that only relates
writes to the same location, and it is the primary technical device
that we use to model what writes can be read. A read, then, can read
from any write executed before it such that the constraints on
coherence order are obeyed.

As one goal of RMC's is to be as weak as possible (but not weaker), the
rules for coherence order are only what are necessary to accomplish
three goals: First, each individual location's should have a total
order of operations that is consistent with program order (this is
\emph{slightly} stronger than just what is required to make
single-threaded programs work.) Second, message passing using
visibility and execution order constraints should work. Third,
read-modify-write operations (like test-and-set and fetch-and-add)
should be appropriately atomic.

\noindent The constraints on coherence order are the following:
\begin{itemize}
\item A read must read from the most recent write
  that it has seen---or from some coherence-later write.
  More precisely, if a read $A$ reads from some write $B$,
  then any other write to
  that location that is \emph{prior to} $A$ must be coherence-before
  $B$.

\item If a write $A$ is \emph{prior to} some other write $B$ to the
  same location, $A$ must be coherence-before $B$.

%% \item
%%   %% If a read $A$ is program-order earlier than some write $B$ to
%%   %% the same location, $B$ must be coherence later than all writes
%%   %% visible to $A$.
%%   If a read $A$ reads from some write $B$, then $B$ must be coherence
%%   earlier than all writes to that location that are later than $A$ in
%%   the program order.
%%   (This disallows reading from writes that are later
%%   in the program order than the read.)
%%   \XXX{This is not how WWP-READ is written,
%%        but I think any edges it doesn't include we can get in other ways}

\item If a read-modify-write operation $A$ reads from a write $B$,
  then $A$ must \emph{immediately} follow $B$ in the coherence order.
  That is, any other write that is coherence-after $A$ must also be
  coherence-after $B$.

\end{itemize}
An action $A$ is \emph{prior to} some action $B$ on the same location
if any of the following holds:
\begin{itemize}
\item $A$ is earlier in program order than $B$.
  (This is crucial for making single threaded programs work properly.)
\item $A$ is \emph{visible to} $B$.
\item $A$ is \emph{prior to} $C$ and $C$ is \emph{prior to} $B$, for some $C$.
\end{itemize}

\noindent
We've discussed \emph{visible to} already, but somewhat more
precisely, $A$ is \emph{visible to} $B$ if:
\begin{itemize}
  \item There is some action $X$ such that $A$ is visibility-ordered
    before $X$ and $X$ is execution-ordered before $B$. That is, there
    is a chain of visibility edges followed by a chain of execution
    edges from $A$ to $B$ (as in the message passing diagrams). Recall
    that reads-from and pushes both induce visibility edges.
\end{itemize}

%\subsection{Non-atomic locations}
%\XXX{do we need to talk about this???}

\section{Compiling RMC}
\label{section:compiler}
RMC is implemented by \texttt{rmc-compiler} \citeComp,
an LLVM plugin which accepts specifications of ordering
constraints and compiles them appropriately. With appropriate support
in language frontends, this allows any language with an LLVM backend
to be extended with RMC support. C and C++ are implemented using macro
libraries that expand to the
specifications expected by \texttt{rmc-compiler}.
%\footnote{Rust has been supported at points as well, but the rapid
%  evolution of Rust nightly has left the code bitrotted}.

%\section{General approach}
A traditional way to describe how to compile language concurrency
constructs is to
give direct mappings from language constructs to sequences of assembly
instructions \cite{batty+:mathematizing, batty+:c++-to-power, sarkar+:synchronizing}.
This approach works well for the C++11 model, in which
the behavior of an atomic operation is primarily determined by
properties of the operation (its memory order, in particular).
In RMC, however, this is not the
case. The permitted behavior of atomic operations (and thus what code
must be generated to implement them) is primarily determined by
the edges between actions. Our descriptions of how to compile RMC
constructs to different architectures, then, focus on edges, and
generally take the form of determining what sort of synchronization
code needs to be inserted \emph{between} two labeled actions with an
edge between them.

\subsection{x86}
%XXX: SPACE: this can be omitted (we've already condensed it)

While it falls short of sequential consistency, x86's memory model
\cite{sewell+:x86-tso} is a pleasure to deal with.
On x86, execution and visibility order come for free---we simply need
to prevent the \emph{compiler} from reordering actions. Pushes can be
implemented with an \texttt{MFENCE} or a locked instruction to some
arbitrary location.

\subsection{ARM and POWER}

Life is not so simple on ARM and POWER, however.
POWER has a substantially weaker memory model
\cite{sarkar+:power} than x86
that incorporates both a very weak memory subsystem in which writes
can propagate to different threads in different orders (that do not
correspond to the order they executed) \emph{and} visible reordering
of instructions and speculation.
For most of our purposes, ARM's model \cite{arm:armv8} is quite similar
to POWER, though writes may not propagate to other threads in
different orders.

Compiling visibility edges is still fairly straightforward. POWER
provides an \texttt{lwsync} (``lightweight sync'') instruction that
does essentially what we need: if a CPU $A$ executes an
\texttt{lwsync}, no write after the \texttt{lwsync} may be propagated
to some other CPU $B$ unless \emph{all} of the writes propagated to
CPU $A$ (including its own) before the
\texttt{lwsync}---the barrier's ``Group A writes'', in the terminology
of POWER/ARM---have also been propagated to CPU $B$. That is, all
writes before
(including those observed from other CPUs) the
\texttt{lwsync} must be visible to another thread before the writes
after it. Then, executing an \texttt{lwsync} between the source and
destination of a visibility edge is sufficient to guarantee visibility
order. The strictly stronger \texttt{sync} instruction on POWER is
also sufficient.
ARM does not have an equivalent to POWER's \texttt{lwsync}, and
so---in the general case---we must use the stronger \texttt{dmb},
which behaves like \texttt{sync}.
ARM does, however, have the \texttt{dmb st} instruction, which
requires that all stores on CPU A before the \texttt{dmb st} become
visible to other CPUs before all stores after the barrier, but imposes
no ordering on loads. This is sufficient for implementing visibility
edges between simple stores.

To implement pushes, we turn to this stronger barrier,
\texttt{sync}. The behavior of \texttt{sync} (and \texttt{dmb}) is
fairly straightforward: the \texttt{sync} does not complete and no
later memory operations can execute until all writes propagated to the
CPU before the \texttt{sync} (the ``Group A writes'') have propagated
to all other CPUs. This is essentially exactly what is needed to
implement a push.

While compiling visibility edges and pushes is fairly straightforward
and does not leave us with many options, compiling execution edges
presents us with many choices to make. ARM and POWER have a number of
features that can restrict the order in which instructions may be
executed:
\begin{itemize}
\item All memory operations prior to a \texttt{sync}/\texttt{lwsync}
  will execute before all operations after it.
\item An \texttt{isync} instruction may not execute until all prior
  branch targets are resolved; that is, until any loads that branches
  are dependent on are executed. Memory operations cannot execute
  until all prior \texttt{isync} instructions are executed.
  % XXX: and addr deps
\item A write may not execute until all prior branch targets are
  resolved; that is, until any loads that the control is dependent on
  are executed.
\item A memory operation can not execute until all reads that the
  address or data of the operation depend on have executed.
\end{itemize}

\noindent
All of this gives a compiler for RMC a bewildering array of options to
take advantage of when compiling execution edges. First, existing
data and control dependencies in the program may already enforce the
execution order we desire, making it unnecessary to emit any
additional code at all. When there is an existing control dependency,
but the constraint is from a read to a read, we can insert an
\texttt{isync} after the branch to keep the order. When dependencies
do not exist, it is often possible to introduce bogus ones: a bogus
branch can easily be added after a read and, somewhat more
disturbingly, the result of a read may be xor'd with itself
(to produce zero) and then added to an address calculation!
And, of
course, we can always use the regular barriers.

This gives us a toolbox
of methods with different characteristics. The barriers, \texttt{sync}
and \texttt{lwsync}, enforce execution order in a many-to-many way:
all prior operations are ordered before all later ones. Using control
dependency is one-to-many: a single read is executed before either all
writes after a branch or all operations after a branch and an
\texttt{isync}. Using data dependencies is one-to-one: the dependee
must execute before the depender.
% PERF IDEAS?
As C++ struggles with finding a variation of the ``consume'' memory
order that compilers are capable of implementing by using existing
data dependencies, we feel that the natural way in which we can take
advantage of existing data dependencies to implement execution edges
is one of our great strengths.

\section{Optimizing RMC Compilation}

\subsection{General Approach}
\label{section:optapproach}

The huge amount of flexibility in compiling RMC edges poses both a
challenge and an opportunity for optimization. As a basic
example, consider compiling the following program for ARM:

\begin{minipage}[b]{0.18\linewidth}
\begin{rmclisting}
VEDGE(wa, wc);
VEDGE(wb, wd);
L(wa, a = 1);
L(wb, b = 2);
L(wc, c = 3);
L(wd, d = 4);
\end{rmclisting}
\end{minipage}

This code has four writes and two edges that overlap with each
other. According to the compilation strategy presented above, to
compile this on ARM we need to place a \ttt{dmb} somewhere between
\li{wa} and \li{wc} and another between \li{wb} and \li{wd}. A naive
implementation that always inserts a \ttt{dmb} immediately before the
destination of a visibility edge would insert \ttt{dmb}s before
\li{wc} and \li{wd}.
A somewhat more clever implementation might insert \ttt{dmb}s greedily
but know how to take advantages of ones already existing---then, after
inserting one before \li{wc}, it would see that the second visibility
edge has been cut as well, and not insert a second \ttt{dmb}.
However, like most greedy algorithms, this is fragile; processing
edges in a different order may lead to a worse solution.
A better implementation would be able to search for places where we
can get more ``bang for our buck'' in terms of inserting barriers.

Things get even more interesting when control flow is in the
mix. Consider these programs:

\begin{minipage}[b]{0.40\linewidth}
\begin{rmclisting}
VEDGE(wa, wb);
L(wa, a = 1);
if (something) {
  L(wb, b = 2);
  // other stuff
}
\end{rmclisting}
\end{minipage}
\begin{minipage}[b]{0.40\linewidth}
\begin{rmclisting}
VEDGE(wa, wb);
L(wa, a = 1);
while (something) {
  L(wb, b = 2);
  // other stuff
}
\end{rmclisting}
\end{minipage}

In both of them, the destination of the edge is executed
conditionally. In the first, it is probably better to insert a barrier
\emph{inside} the conditional, to avoid needing to execute it. The
second, with a loop, is more complicated; which is better depends on
how often the loop is executed, but a good heuristic is probably that
the barrier should be inserted outside of the loop.
% XXX: or ``unroll''

\subsection{Compilation Using SMT}
\label{section:smt}

We model the problem of enforcing the constraints as an SMT problem
and use the Z3 SMT solver \cite{demoura+:z3} to compute the optimal
placement of barriers and use of dependencies (according to our
metrics).
The representation we use was inspired by an
integer-linear-programming representation of graph multi-cut
\cite{costa+:multicut}---we don't go into detail about modeling our
problem as graph multi-cut, since it is no more
illuminating than the SMT representation and does not scale up to
using dependencies. This origin survives in our use of the word ``cut''
to mean satisfying a constraint edge.

Compilation proceeds a function at a
time. Given the set of labeled actions and constraint edges and the
control flow graph for a function, we produce an SMT problem with
solutions that indicate where to insert barriers and where to take
advantage of (or insert new) dependencies. The SMT problem that we
generate is \emph{mostly} just a SAT problem, except that integers are
used to compute a cost function, which is then minimized.
%Z3 has built in support for minimizing a quantity.
%but even without that, the cost can be minimized by adding a
%constraint that bounds it, and then binary searching on that bound.

When considering the function's CFG, we assume that each labeled
action lives in a basic block by itself. Furthermore,
we extend the
CFG to contain edges from all exits of the function to the entrance of
the function, in order to model paths into future invocations of the
function.

As a preprocessing step, we compute the transitive closure of all of
the constraint edges for the function (taking into account that
visibility implies execution). We can then safely
ignore any edges that don't have any meaning apart from their
transitive effects, such as those involving no-ops.

We present two versions of this problem. As an introduction, we first
present a complete system that always uses barriers, even when
compiling execution edges.  We then discuss how to generalize it to
use control and data dependencies.

\subsubsection{Barrier-only implementation}

\newcommand{\mand}[1]{\bigwedge_{#1}}
\newcommand{\mor}[1]{\bigvee_{#1}}
\newcommand{\svar}[1]{{\sf #1}}
\newcommand{\ivar}[1]{#1}
\newcommand{\uvar}[1]{{\sf #1}}
\newcommand{\ovar}[1]{{\overline{\sf #1}}}

The rules for encoding the compilation of visibility edges as an SMT
problem are reasonably straightforward
(for space and simplicity we omit the potential use of \texttt{dmb st}):
\begin{eqnarray*}
&& \mand{s \vo t} \svar{vcut}(s, t) \\
\svar{vcut}(s, t) &=& \mand{p \in {\sf paths}(s, t)} \svar{vcut\_path}(p) \\
\svar{vcut\_path}(p) &=& \mor{e \in p} \ovar{lwsync}(e) \lor \ovar{sync}(e) \\
\end{eqnarray*}
We write $\svar{foo}(x)$ to mean a variable in the SMT problem that is
given a definition by our equations and $\ovar{foo}(x)$ to mean an
``output'' variable whose value will be used to drive compilation.
Later in this section, we will use $\ivar{foo}(x)$ to mean an
``input'' variable that is not a true SMT variable at all, but a
constant set by the compiler based on some analysis of the program.

Here, the assignments to the $\ovar{lwsync}$ and $\ovar{sync}$
variables produced by the SMT solver are used by the compiler to
determine where to insert barriers.
We write $s \vo t$ to quantify over visibility edges from
$s$ to $t$ and take ${\sf paths}(s, t)$ to
mean all of the simple paths from $s$ to $t$.
Knowing that, these rules state that (1) every visibility edge must be
cut, (2) that to cut a visibility edge, each path between the source
and sink must be cut, and (3) that to have a visibility cut on a path
means deciding to insert \ttt{sync} or \ttt{lwsync} at one of the
edges along the path.

Since in the version we are presenting now, we only use barriers to
enforce execution order, the condition for an execution edge is the
same as that for a visibility one (writing $s \xo t$ to quantify over
execution edges):
\begin{eqnarray*}
&& \mand{s \xo t} \svar{vcut}(s, t) \\
\end{eqnarray*}

The rules for compiling push edges are straightforward: they are
essentially the same as for visibility, except only heavyweight syncs
are sufficient to cut an edge (writing $s \pushe t$ to quantify over
push edges):
\begin{eqnarray*}
&& \mand{s \pushe t} \svar{pcut}(s, t) \\
\svar{pcut}(s, t) &=& \mand{p \in {\sf paths}(s, t)} \svar{pcut\_path}(p) \\
\svar{pcut\_path}(p) &=& \mor{e \in p} \ovar{sync}(e) \\
\end{eqnarray*}

All of the rules shown so far allow to find \emph{a} set of places to
insert barriers, but we could have done that already without much
trouble. We want to be able to \emph{optimize} the placement. This is
done by minimizing the following quantity:
\[
  \sum_{e \in E} \ovar{lwsync}(e) w(e) cost_{\sf lwsync} +
  \ovar{sync}(e) w(e) cost_{\sf sync}
\]
Here, we treat the boolean variable representing barrier insertions
as $1$ if they are true and $0$ if false. The $w(e)$ terms represent
the ``cost'' of an edge---these are precomputed based on how many
control flow paths travel through the edge and whether it is inside of
loops. The $cost_{\sf lwsync}$ and $cost_{\sf sync}$ terms are weights
representing the costs of the \ttt{lwsync} and \ttt{sync}
instructions, and should be based on their relative costs.

\subsubsection{Dependency trickiness}

The one major subtlety that needs to be handled when using
dependencies to enforce execution ordering is that ordering must be
established with \emph{all} subsequent occurrences of the
destination. Consider the following code:

\begin{minipage}[b]{0.76\linewidth}
\begin{rmclisting}
rmc::atomic<int> x, y;
void f(bool b) {
  XEDGE(ra, wb);
  int i = L(ra, x);
  if (b) return;
  if (i == 0) {
    L(wb, y = 1);
  }
}
\end{rmclisting}
\end{minipage}

In this code, we have an execution edge from \li{ra} to \li{wb}. We
also have a control dependency from \li{ra} to \li{wb}, which we may
want to use to enforce this ordering. There is a catch,
however---while the execution of \li{wb} is always control dependent
on the result of the \li{ra} execution from the current invocation of
the function, it is \emph{not} necessarily control dependent on
executions of \li{ra} from previous invocations of the function (which
may have exited after the conditional on \li{b}).

The takeaway here is that we must be careful to ensure that the
ordering applies to all future actions, not just the
closest. Just because an action is dependent on
a load does not mean it is necessarily dependent on all prior
invocations of the load. Our solution to this is, when using a
dependency to order some actions $A$ and $B$, to additionally require
that $A$ be execution ordered with subsequent invocations of itself.
If we are are using a control
dependency to order $A$ and $B$, we can get a little weaker---it
suffices for future executions of $A$ to be control dependent on $A$,
even if that would not be enough to ensure execution order on its
own.

%% Of course, we can't directly enumerate all possible paths
%% between the actions, since there are infinitely many.
%% \XXX{A BUNCH OF THIS IS WRONG}
%% But we \emph{do}
%% consider all simple paths---this means that any path from an action
%% $A$ to $B$ that we do not consider must proceed from $A$ to $A$ again,
%% and then to $B$. We take advantage of this when using dependencies to
%% ensure ordering: if we can use dependencies to order $A$ along all
%% simple paths to $B$, and can also execution order $A$ with $A$, then
%% this suffices to order $A$ and $B$.

\subsubsection{Supporting dependencies}

With this in mind, we can now give the constraints that we use for
handling execution order. They are considerably more hairy than those
just using barriers. First, the ``top-level rules'':

\begin{eqnarray*}
&& \mand{s \xo t} \svar{xcut}(s, t) \\
\svar{xcut}(s, t) &=& \mand{p \in {\sf paths}(s, t)} \svar{xcut\_path}(p) \\
\svar{xcut\_path}(p) &=& \svar{vcut\_path}(p) \lor \\
&& (\svar{ctrlcut\_path}(p) \land \\
&& \,\,(\svar{ctrl}(s, s) \lor \svar{xcut}(s, s))) \lor \\
&& (\svar{datacut\_path}(p) \land \svar{xcut}(s, s)) \\
&& \text{(where $s = {\sf head}(p)$)}
\end{eqnarray*}
As discussed above, execution order edges can be cut along a path by
barriers as with visibility (\svar{vcut\_path}), and also by control
(\svar{ctrlcut\_path}) and data (\svar{datacut\_path})
dependencies, if the appropriate side conditions hold.

Then, the rules for cutting edges using control (for
simplicity, and because they don't help much on ARM, we leave out the
rules for using \texttt{isync}):
\begin{eqnarray*}
\svar{ctrl}(s, t) &=& \mand{p \in {\sf paths}(s, t)} \svar{ctrl\_path}(p) \\
\svar{ctrl\_path}(p) &=& \mor{e \in p} \ivar{can\_ctrl}(s, e) \land \ovar{use\_ctrl}(s, e)\\
&& \text{(where $s = {\sf head}(p)$)} \\
\svar{ctrlcut\_path}(p) &=&
\ivar{iswrite}(t) \land \svar{ctrl\_path}(p) \\
&& \text{(where $t = {\sf tail}(p)$)} %%\\
%% \svar{ctrlisync\_path}(p) &=& \mor{e::p' \in p}
%%   \svar{ctrl\_path}(s, p) \land \svar{isync\_path}(p')\\
%% && \text{(where $s = {\sf head}(p)$)} \\
%% \svar{isync\_path}(p) &=& \mor{e \in p} \svar{isync}(e)
\end{eqnarray*}
In this, $\ivar{can\_ctrl}(s, e)$ is
true if there is---or it would be possible to add---a branch along the
edge $e$ that is dependent on the value read in
$s$ and
$\ivar{iswrite}(t)$ is true if $t$ is an
action containing only writes.
%% In the notation $e::p' \in p$, we
Then $\ovar{use\_ctrl}(s, e)$ is the corresponding output,
indicating whether the branch will be for ordering purposes.
%% intend for $e$ to represent an edge in $p$ and $p'$ to represent the
%% rest of the path after $e$, starting with $e$'s destination.

Here, we can cut an execution edge along a path using control
dependencies if the destination of the edge is a write and somewhere
along the path there is a branch that is dependent on the source
value.
This could be extended to support using \ttt{isync} regardless of what
sort of action the destination is by requiring an \ttt{isync} along
the path after the branch.

Data dependency is quite simple on the SMT side of things:
%% \begin{eqnarray*}
%% \svar{datacut\_path}(p) &=& \mor{(\_,t)::p' \in p}
%% %  \svar{data}(s, t, p) \land (\svar{ctrlcut\_path}(p') \lor \svar{datacut\_path}(p'))   \\
%%   \svar{data}(s, t, p) \land \svar{cut\_path}(p') \\
%% && \text{(where $s = {\sf head}(p)$)} \\
%% \svar{cut\_path}(p) &=& \svar{ctrlcut\_path}(p') \lor \svar{datacut\_path}(p')  \\
%% \svar{data}(s, t, p) &=& \svar{can\_data}(s, t, p) \land \svar{use\_data}(s, t, p)
%% \end{eqnarray*}
\begin{eqnarray*}
\svar{datacut\_path}(p) &=& \ivar{can\_data}(s, t, p) \land \ovar{use\_data}(s, t, p) \\
&& \text{(where $s = {\sf head}(p), t = {\sf tail}(p)$)}
\end{eqnarray*}
%% In the notation $e::p' \in p$, we
%% intend for $e$ to represent an edge in $p$ and $p'$ to represent the
%% rest of the path after $e$, starting with $e$'s destination.
Here, $\ivar{can\_data}(s, v, p)$ is
true if there is
a data dependency from $s$ to $v$, following the path
$p$ (it could also be extended to mean that a dependency could be
\emph{added}, but the compiler does not currently do that).
The path
$p$ needs to be included because whether something is data-dependent
can be path-dependent (in LLVM's SSA based intermediate
representation, this idea is made explicit through the use of phi
nodes).

This check is somewhat subtle: the paths we are trying to cut
are only simple paths, but actual execution can follow complex paths
(ones with cycles). Thus, we actually must check that a
data dependency exists not just when following the simple path $p$,
but also when following any path that ``detours'' away from $p$ and
then returns to it. The data dependence check, then, works by tracing
backwards the chain of instruction uses from the destination, looking
for the source. When an operand in this chain could come from multiple
different basic blocks (via an SSA phi node), we do the check for every
source that could have been reached while executing along path $p$
while possibly taking detours off of it.
% XXX: MEH
%XXX: discuss the algorithm? certainly in the thesis!
%While an \textt{lwsync} or a control-dependency

%% Given this, though, the idea here is straightforward: a constraint is
%% cut by data dependencies if there is a chain of data deps---possibly
%% concluded by a control dep---from the source to the sink. (Note that
%% $\svar{datacut\_path}(p')$ is vacuously true if the path is empty,
%% providing a base case.)

The only thing that remains is to extend the cost function to take
into account how we use dependencies. This proceeds by giving weights
to $\ovar{use\_data}$ and $\ovar{use\_ctrl}$ and
summing them up. Different weights should be given based on whether
the dependencies are already present or need to be synthesized.
Currently we use values for the weights that we find work well in
practice, though they have not yet been carefully optimized or derived
in any particularly principled way.

\subsubsection{Scoped constraints}
The one major thing lacking from the rules as presented so far is any
consideration of \li{VEDGE_HERE} and \li{XEDGE_HERE}. Extending our
system to handle scoped constraints is relatively
straightforward. Recall that a scoped constraint between $a$ and $b$
establishes edges between executions of $a$ and $b$, but only when the
edges do not leave the scope of the constraint. Since we define the
scope of a constraint in terms of the control flow graph, this means
that $a$ and $b$ must be appropriately ordered along all control flow
paths that do not pass through the binding site of the constraint.

With that in mind, the extensions to the rules are simple. For
visibility (and push) edges, it is a simple matter of only requiring
that we cut paths not containing the binding site:
\begin{eqnarray*}
&& \mand{s \vo t @ b} \svar{vcut}(b, s, t) \\
%&& \mand{s \pushe t @ b} \svar{pcut}(b, s, t) \\
\svar{vcut}(b, s, t) &=& \mand{p \in {\sf paths\_wo}(b, s, t)} \svar{vcut\_path}(p) \\
%\svar{pcut}(b, s, t) &=& \mand{p \in {\sf paths\_wo}(b, s, t)} \svar{pcut\_path}(p)
\end{eqnarray*}
Here we write $s \arr{} t @ b$ to indicate a constraint edge from $s$
to $t$ that is bound at $b$ and ${\sf paths\_wo}(b, s, t)$ to mean all
simple paths from $s$ to $t$ without $b$ in them. Non-scoped
constraints will have a dummy $b$ that does not appear in the CFG.

The modifications for execution edges are similar but have one
additional wrinkle: when using control and data dependencies to ensure
ordering, $\svar{xcut\_path}$ can appeal to $\svar{xcut}$ and
$\svar{ctrl}$; we modify $\svar{xcut\_path}$ to pass the binding site
down to these. (In fact, this is the main use-case of scoped
constraints: eliminating the need to order successive invocations when
using data dependencies.) Since it can affect whether a data
dependency exists along all path detours, we must also add a binding
site argument to $\svar{datacut\_path}$ that is passed down into
$\ivar{can\_data}$ and $\ovar{use\_data}$.

\begin{eqnarray*}
&& \mand{s \xo t @ b} \svar{xcut}(b, s, t) \\
\svar{xcut}(b, s, t) &=& \mand{p \in {\sf paths\_wo}(b, s, t)} \svar{xcut\_path}(b,p) \\
\svar{xcut\_path}(b, p) &=& \svar{vcut\_path}(p) \lor \\
&& (\svar{ctrlcut\_path}(p) \land \\
&& \,\,(\svar{ctrl}(b, s, s) \lor \svar{xcut}(b, s, s))) \lor \\
&& (\svar{datacut\_path}(b, p) \land \svar{xcut}(b, s, s)) \\
&& \text{(where $s = {\sf head}(p)$)} \\
%\end{eqnarray*}
%\begin{eqnarray*}
\svar{ctrl}(b, s, t) &=& \mand{p \in {\sf paths\_wo}(b, s, t)} \svar{ctrl\_path}(p) \\
\svar{datacut\_path}(b, p) &=& \ivar{can\_data}(b, s, t, p) \land \ovar{use\_data}(b, s, t, p) \\
&& \text{(where $s = {\sf head}(p), t = {\sf tail}(p)$)}
\end{eqnarray*}

\subsection{Using the solution}

While the process of using the SMT solution to insert barriers and
take advantage of dependencies is fairly straightforward, there are a
handful of interesting subtleties.

% XXX: SPACE: could cut a lot of this
The first snag is that while our SMT problem thinks in terms of
inserting barriers at control flow \emph{edges}, we actually have to
insert the barriers into the inside of basic blocks. This presents a
snag when we are presented with control flow graphs like this:
\begin{center}
\incggen[width=4cm]{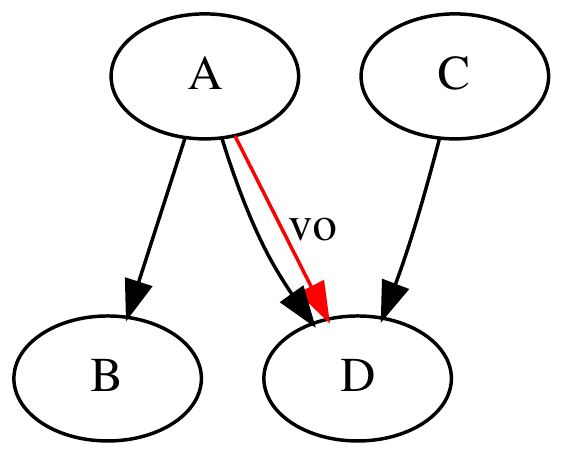}
\end{center}
The SMT solver will ask for a barrier to be inserted between basic
blocks A and D, but there is a catch:
the edge is a \emph{critical edge} (a CFG edge where the source has
multiple successors and the destination has multiple predecessors)
for which there is no way to insert
code along it without the code running on other paths as well.
Fortunately, this is a well-known issue, so LLVM has a pass for
breaking critical edges by inserting intermediate empty basic
blocks. By requiring critical edges to be broken,
we can always safely insert the barrier at
either the end of the source or the start of the destination.
%If placed at the end of
%A, it runs even when jumping to B. If placed at the beginning of D, it
%runs even when control came from C.
%x
This sort of barrier placement, which falls out naturally in our
implementation of RMC, can't be achieved using C++11 memory orders on
operations (though it could be achieved by manually placing C++'s low
level thread fences).

The other snag comes when taking advantage of data and control
dependencies: we need to make sure that later optimization phases can
not remove the dependency. This ends up being a somewhat grungy
engineering problem. Our current approach involves disguising the
provenance of values involved in dependency chains to ensure that
later passes never have enough information to perform breaking
optimizations.

\subsection{ARMv8}
ARMv8 brings with it a number of additions to the ARM memory model
that we can take advantage of \cite{arm:armv8}.

ARMv8 adds the \ttt{LDA} and \ttt{STL} families of instructions, which
they name ``Load-Acquire'' and ``Store-Release''.  Designed to
efficient implement C++11 SC atomics, these instructions may be used
to realize execution and visibility edges.  In RMC terms,
Store-Release writes become visible after all program-order prior
stores and all stores observed by program-order prior loads, and so
are visibility-after all program order predecessors.  Load-Acquire
reads, on the other hand, execute before all program-order successors.

ARMv8 introduces a new weaker variant of \texttt{dmb}, written
\texttt{dmb ld}. The \texttt{dmb ld} barrier orders
all loads before the barrier before all stores
and loads after it. Because all meaningful execution edges have a load
as their source, this means that \texttt{dmb ld} can satisfy any
execution edge with a straightforward barrier weaker than that needed
for visibility, thus cleanly filling a niche that was left empty on
POWER and ARMv7.

%% Despite these names, they are
%% actually strong enough to implement C++11's SC atomics \cite{cam:mappings}:
%% ``Store-Releases'' become visible before any subsequent
%% ``Load-Acquires'' execute.

% XXX: space
%As of the the latest ARM ARM,
ARM is ``Other-multi-copy atomic''---informally, if a write is
observed by some processor \emph{other} than the one who performed it,
then it is observed by \emph{all} other processors.  One outcome of
this is that \texttt{dmb ld} can \emph{almost} be used to implement
visibility edges from loads to stores: if a load reading from another
thread's write is made to execute before a store, Other-multi-copy
atomicity ensures that the two writes are observed in the correct
order by all processors. But this does not hold if the load reads from
a store done by the \emph{same} processor!
This snag suggests a workaround: adding a \texttt{dmb st} to ensure
that any earlier same-processor stores must become visible before
subsequent ones.
This means that on ARMv8, \texttt{dmb ld; dmb st} can serve
essentially the same role as Power's \texttt{lwsync}: cutting
arbitrary visibility edges while being short of a full fence.  Perhaps
surprisingly, taking advantage of this actually yields performance
wins!

\section{Evaluation}
\label{section:eval}

\begin{figure}[]
  \hspace{-2em}
%\begin{minipage}[b]{0.5\linewidth}
\incggen[width=0.9\linewidth]{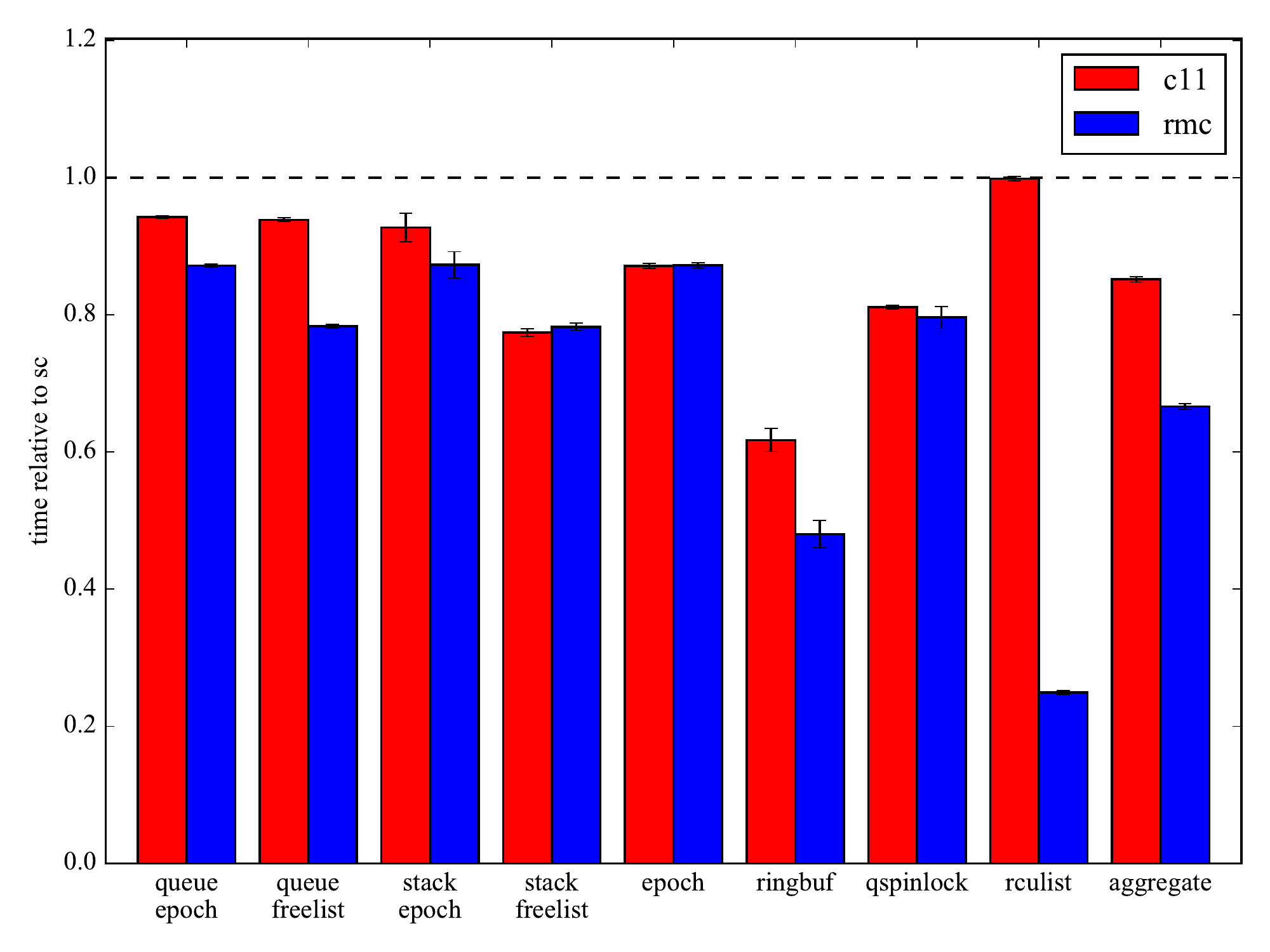}
\caption{ARMv7 benchmarks}
\label{fig:armv7_bench}
\end{figure}

\begin{figure}[]
\incggen[width=0.9\linewidth]{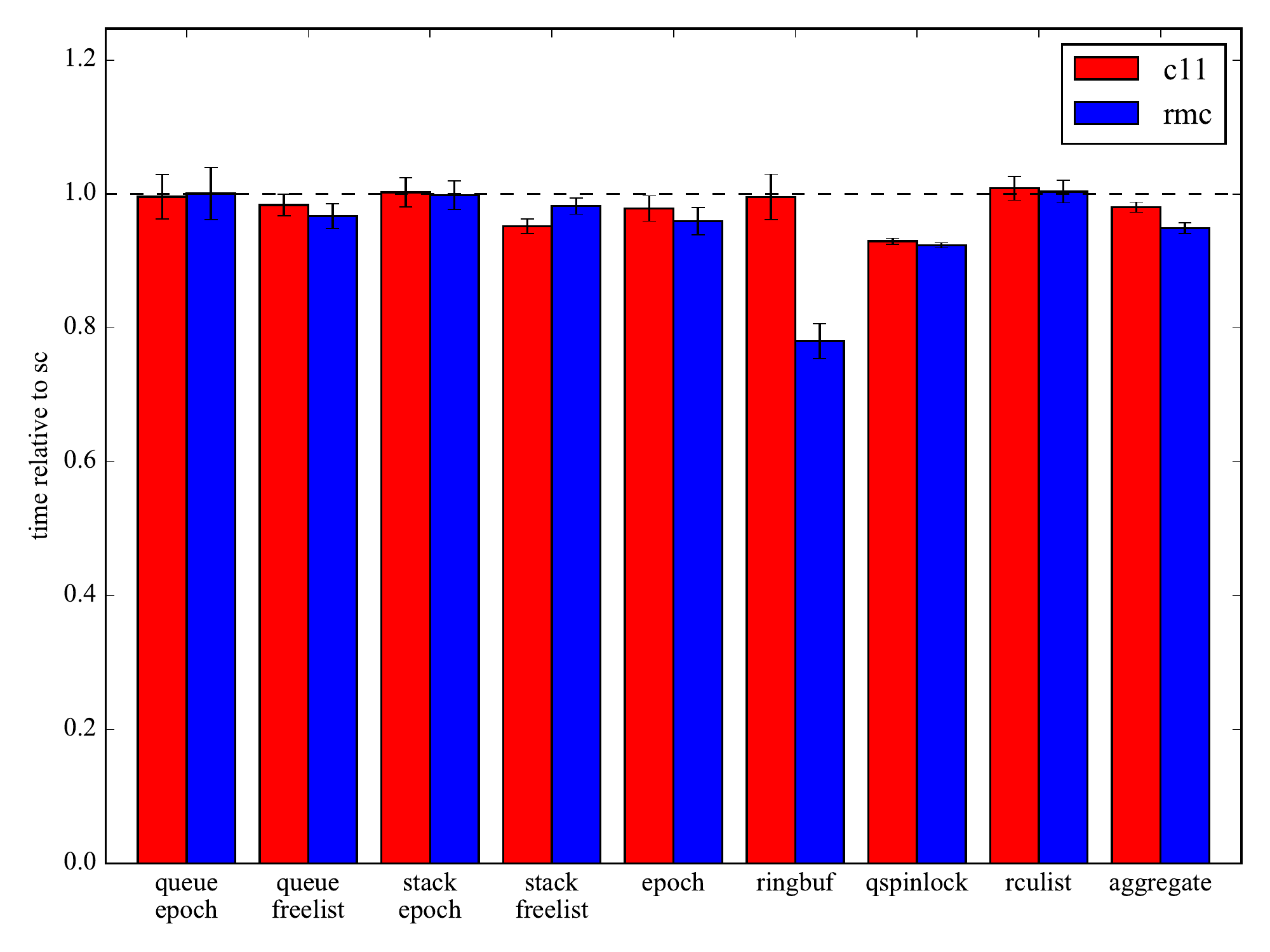}
\caption{ARMv8 benchmarks}
\label{fig:armv8_bench}
%\end{minipage}
\end{figure}

\begin{figure}[]
\incggen[width=0.9\linewidth]{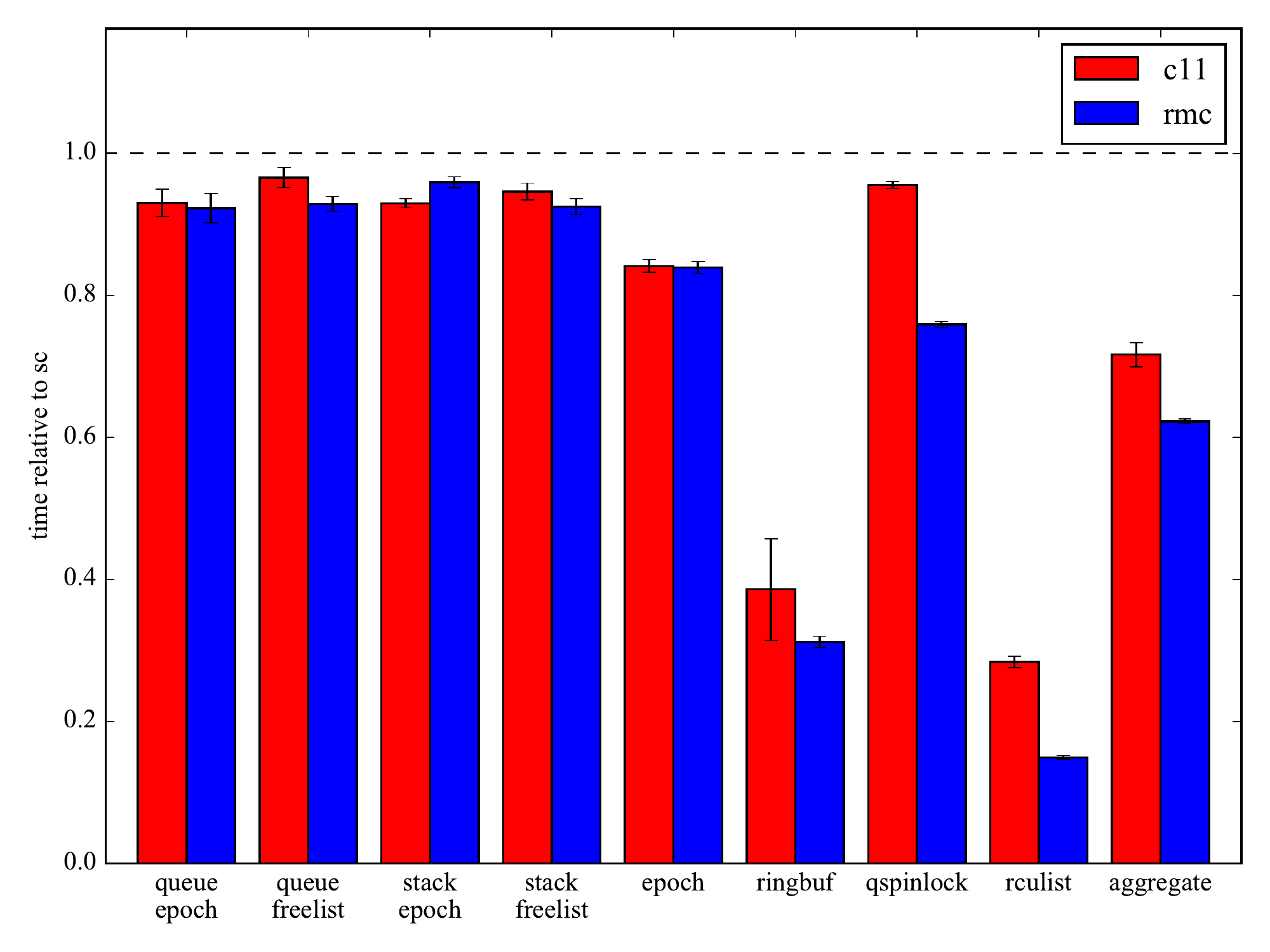}
\caption{Power benchmarks}
\label{fig:power_bench}
%\end{minipage}
\end{figure}

To evaluate \texttt{rmc-compiler}, we implemented a number of
low-level concurrent data structures using C++ SC atomics, C++
low-level atomics, and RMC and measured their performance on
ARMv7 (Figure \ref{fig:armv7_bench}),
ARMv8 (Figure \ref{fig:armv8_bench}),
and Power (Figure \ref{fig:power_bench}).
The graphs plot the performance of the C++ low-level atomic and RMC
versions relative to that of the SC version. The ``aggregate'' column
shows the geometric mean of the other tests.
We performed our ARMv7 tests on an NVIDIA Jetson TK1 quad-core
ARM Cortex-A15 board and our ARMv8 tests on an
ODROID-C2 quad-core ARM Cortex-A53 board.
Power tests were performed on an IBM Power System E880 (9119-MHE).
Tests were compiled using Clang/{\mbox LLVM} 3.9.

%\end{minipage}
%\end{figure}
%\begin{figure}[]
%\hspace{-2em}
%\begin{minipage}[b]{0.5\linewidth}

%\begin{figure}[]
%\hspace{-2em}
%\begin{minipage}[b]{0.5\linewidth}

%% In Figure \ref{fig:ms_queue} we show the results of testing a number
%% of implementations of Michael-Scott concurrent queues
%% \cite{michael+:queues}.
The ``queue'' and ``stack'' tests are implementations of Michael-Scott
concurrent queues \cite{michael+:queues} and Treiber stacks
\cite{treiber:coping} and each come in two varieties.
The ``freelist'' versions are traditional and use generation counts
and Treiber-stack-based free lists for memory management
while the ``epoch'' versions use an elegant memory management
technique called epoch-based reclamation
\cite{fraser:lockfreedom} that allows threads to register objects for
freeing once all threads are done using them.
\footnote{Epoch reclamation is essentially a variant of RCU with an
  increased focus on efficient memory reuse.}
The ``epoch'' tests measure the
performance of different versions of the epoch library while running
stack and queue tests.
The ``ringbuf'' test is a ring-buffer very similar to that discussed
in Section \label{section:ringbuf}. The ``qspinlock'' test is an
implementation of a queue-based spinlock scheme used in the
Linux kernel \cite{corbet:qspinlocks}.
%based on prior work called ``MCS locks'' \cite{mellor+:mcslocks}.

%In Figure \ref{fig:rculist} we show the results of testing operations
%on a list protected by RCU.
The ``rculist'' test measures operations on an RCU-protected linked
list.
Since RCU is optimized for read-mostly
data structures, the test interleaves lookups of list
elements and list modifications with one modification every 10000
lookups.
Here, list modifications may occur while readers are
traversing the list, so reader threads must take care to ensure that
they actually see the contents of any node they get a pointer to. This
turns out to be a perfect case for taking advantage of existing data
dependencies to enforce ordering.
The C++11 version uses \li{memory_order_consume}, which
establishes ordering based on data dependencies. Unfortunately,
consume remains not-implemented-as-intended on all major compilers,
which simply emit barriers, leading to uninspiring performance.
The RMC version uses fine-grained execution orderings established
using \li{XEDGE_HERE} and successfully avoids emitting any barriers on
the read side.
%
%% The ``linux'' test is the C++11 version with all uses
%% of consume replaced with relaxed memory accesses;
%% this, following the standard practice in the Linux kernel, confronts
%% the lack of assurances from the language by instead attempting to
%% quantify over all possible compiler transformations
%% \cite{mckenney:rcu-dereference}. While this approach is fraught with
%% danger, it has served extremely well from a performance
%% perspective. We consider it a major success that we can match its
%% performance while actually providing a well-defined semantics!

Overall, RMC shows a modest performance win on all three tested
architectures. On all three, RMC shows modest wins on most (though not
all) data structure tests. RMC gets solid wins on the ringbuf test due
to its ability ot take advantage of a control dependency to enforce
execution order. The epoch library tests show little difference, which
makes sense, as C++11 and RMC should generate essentially identical
code in the fast path common case. In RCU-protected list manipulation,
RMC wins tremendously on ARMv7 and Power by virtue of being able to
rely on data dependencies for enforcing execution order. On ARMv8,
however, this yields no real speedup over using Load-Acquire
instructions!
%XXX: say more about ARMv8
Though not shown in this chart, RMC matches the performance of RCU
list search done in the Linux kernel style of
attempting to quantify over all possible compiler transformations
so as to avoid doing anything that may result in a dependency being
optimized away
\cite{mckenney:rcu-dereference}.
While this approach is fraught with
danger, it has served extremely well from a performance
perspective. We consider it a major success that we can match its
performance while actually providing a well-defined semantics!

%% In Figure \ref{fig:seqlock} we show the results of testing operations on a seqlock
%% \cite{hemminger:frlock, boehm:seqlocks}.

%% In Figure \ref{fig:epoch}, we investigate the performance of different
%% implementations of the epoch memory reclamation library while using it
%% for queues, stacks, and RCU protected lists. For each data structure,
%% we hold the implementation of the structure constant while varying the
%% library: stacks and queues use the ``c11'' version while RCU-protected
%% lists use the ``linux'' version. We test two versions using C++11
%% relaxed atomics: in ``c11simp'', the common-case hot-path (that is
%% executed for every operation on epoch protected data structures) is
%% implemented with relaxed memory orders but the slow path for trying to
%% collect memory uses more straightforward SC atomics. In ``c11'', the
%% entire library uses weak memory orderings \footnote{Parts of the
%%   ``c11'' version were designed by essentially running the RMC
%%   placement algorithm in our heads based on the RMC
%%   version}. Unsurprisingly, it turns out not to matter much: all three
%% tested epoch implementations that use weak orderings perform at about
%% the same level for most workloads; they all generate essentially the
%% same code for the small fast-path code and any differences in the more
%% complex slow-path are drowned out by its infrequency. Only testing RCU
%% lists with an anomalously write-heavy load was able to suss out any
%% performance differences.

\subsection{Compiler performance}
Given the rather high asymptotic complexity of our SMT-based
algorithms
(the size of the generated SMT problems is linear in the
number of paths through the control flow graph between the sources and
destinations of edges, which can be exponential in the worst case),
it is natural to wonder about the performance of the compiler itself.
%
%In practice, \texttt{rmc-compiler} performs quite acceptably.
Across our test suit, the (geometric) average slowdown was
{\integrityaggregateSlowdown}x on ARMv7,
{\weaknessaggregateSlowdown}x on ARMv8,
and {\yeezyaggregateSlowdown}x on POWER---
the difference across architectures is due to having more options for
implementing constraints on ARMv8 and POWER.
The slowdown varies based on the complexity of the test case---RCU
list searching on POWER has a slowdown of {\yeezyrculistSlowdown}x
while the very complex qspinlock test has a slowdown of
{\yeezyqspinlockSlowdown}x.

There are a number of approaches that could be taken
to improve compilation time, including attempting to reduce the SMT
state space by eliminating unlikely options, adding RMC-tuned
heuristics to Z3, and specifying a timeout to Z3 and choosing the best
found.
We are not particularly troubled by the compilation times, however, as
the sort of low-level concurrent code that uses RMC is likely to be a
relatively small portion of any real codebase.

%% While we we
%% have not conducted a rigorous study of compiler performance, we have
%% also never been frustrated with its performance during testing. As one
%% data point: compiling and linking the queue benchmark and epoch
%% library on our Jetson TK1 ARM test machine took 6.0s for the RMC
%% version and 5.4s for the C++11 version.

\section{Related Work}
Using integer linear programming (ILP) to optimize the placement of barriers
is a common technique. \citet{bouajjani+:enforcing-against-tso} and
\citet{alglave+:musketeer} both use ILP to calculate where to
insert barriers to recover sequential consistency as part of tools for
that purpose.
\citet{bender+:declarative-fence} use ILP to place
barriers in order to compile what is essentially a much simplified version
of RMC (with only one sort of edge, most akin to RMC's push
edges). We believe we are the first to extend this technique to
handle the use of dependencies for ordering.

%% \citeauthor{ou+:automo}'s AutoMO \citey{ou+:automo} attempts to simplify the use
%% of the C++11 model by automatically inferring memory orders: given a
%% data structure implementation and a test suite, AutoMO finds a
%% assignment of memory orders such that the data structure exhibits only
%% SC behaviors under the test suite.

OpenMP's \cite{openmp:openmp} flush directive also allows a form of
pairwise ordering, but the programming model has little in common with
RMC: OpenMP flush is just a barrier---with restrictions on what
locations it applies to---and the pairwise ordering is based on
locations and not specific actions.

There are proposals by \citet{mckenney+:new_consume,  mckenney+:marking_consume}
to change the semantics of consume to be more realistically
implementable, but nothing seems finalized yet .

%\appendix
%\section{Appendix Title}
%This is the text of the appendix, if you need one.

\begin{acks}

Joseph Tassarotti provided helpful feedback and input.
NVIDIA Corporation donated a Jetson TK1 development
kit which was used for testing and benchmarking.
The IBM Power Systems Academic Initiative team provided access to a
POWER8 machine which was used for testing and benchmarking.

\end{acks}

% We recommend abbrvnat bibliography style.
\clearpage
\bibliography{citations}{}

\end{document}